\documentclass[twocolumn,secnumarabic,amssymb, nobibnotes, aps, prd, superscriptaddress]{revtex4}

\usepackage[dvips]{graphicx}

\begin{document}

\title{Photodissociation as a probe of the H$_3^+$ avoided crossing seam}

\author{Xavier Urbain }
\email{xavier.urbain@uclouvain.be}
\affiliation{Institute of Condensed Matter and Nanosciences, Universit\'e catholique de Louvain, 2 Chemin du Cyclotron, 1348, Louvain-la-Neuve, Belgium}
\author{Arnaud Dochain}
\affiliation{Institute of Condensed Matter and Nanosciences, Universit\'e catholique de Louvain, 2 Chemin du Cyclotron, 1348, Louvain-la-Neuve, Belgium}
\author{Rapha\"el Marion}
\affiliation{Institute of Condensed Matter and Nanosciences, Universit\'e catholique de Louvain, 2 Chemin du Cyclotron, 1348, Louvain-la-Neuve, Belgium}
\author{Thibaut Launoy}
\affiliation{Institute of Condensed Matter and Nanosciences, Universit\'e catholique de Louvain, 2 Chemin du Cyclotron, 1348, Louvain-la-Neuve, Belgium}
\affiliation{Laboratoire de Chimie Quantique et Photophysique, Universit\'e libre de Bruxelles, 50 av. F.D. Roosevelt, CP160/09, 1050 Brussels, Belgium}
\author{J\'er\^ome Loreau }
\affiliation{Institute of Condensed Matter and Nanosciences, Universit\'e catholique de Louvain, 2 Chemin du Cyclotron, 1348, Louvain-la-Neuve, Belgium}
\affiliation{Laboratoire de Chimie Quantique et Photophysique, Universit\'e libre de Bruxelles, 50 av. F.D. Roosevelt, CP160/09, 1050 Brussels, Belgium}

\begin{abstract}
{Experiments are conducted to investigate the role of the avoided crossing seam in the photodissociation of H$_3^+$. Three-dimensional imaging of dissociation products is used to determine the kinetic energy release and branching ratio among the fragmentation channels. Vibrational distributions are measured by dissociative charge transfer of H$_2^+$ products.  It is found that the photodissociation of hot H$_3^+$ in the near ultraviolet produces cold H$_2^+$, but hot H$_2$.  Modelling the wavepacket dynamics along the repulsive potential energy surface accounts for the repopulation of the ground potential energy surface. The role of the avoided crossing seam is emphasized and its importance for the astrophysically relevant charge transfer reactions is underlined.}

\end{abstract}

\maketitle

\section{Introduction}

The molecular ion H$_3^+$ plays a pivotal role in many astrophysical environments \cite{Tennyson1995,Oka2013}.
In the interstellar medium, H$_3^+$ acts as a proton donor and provides the main path towards the formation of hydrides through the reaction H$_3^+$ + M $\rightarrow$ H$_2$ + MH$^+$, which in turn leads to a sequence of ion-neutral reactions.
H$_3^+$ is also a key species to understand deuterium fractionation: the main reservoir of deuterium is HD, which reacts with H$_3^+$ to form H$_2$D$^+$. This leads to rapid deuteration at low temperature, as the reverse reaction is endothermic by 139.5 K. H$_2$D$^+$ then efficiently transfers its D in ion-neutral reactions, the differences in zero-point energy with respect to H-containing species leading to extreme deuteration at temperatures below 10-20 K. 

The formation of H$_3^+$ in binary collisions of H$_2$ with H$_2^+$ is controlled by the ionization of H$_2$. While cosmic rays are the main source of ionization in the interstellar medium, the reaction of vibrationally excited H$_2$ with protons is at play in hotter regions like shocked molecular clouds and the atmosphere of giant planets. Collision-induced vibrational excitation of H$_2$ triggers the charge transfer reaction: 
\begin{equation} \label{H2H+}
\mathrm{H}_2(v\geq 4) + \mathrm{H}^+ \rightarrow \mathrm{H}_2^+(v^+) + \mathrm{H}.
\end{equation}
In diffuse clouds where the density of H$_2$ is significantly lower than in dense clouds, the reverse reaction competes with the formation of H$_3^+$ from H$_2^+$:
\begin{equation} \label{H2+H}
\mathrm{H}_2^+ + \mathrm{H} \rightarrow \mathrm{H}_2(v) + \mathrm{H}^+.
\end{equation}
Both of these reactions are mediated by the H$_3^+$ potential energy surface. 
Numerous measurements of the cross section of reaction (\ref{H2H+}) exist, albeit with H$_2$ in its vibrational ground state and at collision energies well above 10 eV \cite{Kusakabe2004}. Our recent work on charge transfer in proton H$_2$ collisions  down to 15 eV \cite{Urbain2013} demonstrated the decisive role of the avoided crossing seam connecting the ground and first excited potential energy surfaces of H$_3^+$. The vibrational population of the H$_2^+$ products was shown to peak at $v^+=0$ at a collision energy of 45 eV, evolving from a Franck-Condon distribution at keV energies. The current understanding is that the proton approach triggers the vibrational excitation necessary for the reaction (\ref{H2H+}) to proceed. This resonant population of $v^+=0$ vanishes at even lower energies.

The reverse reaction (\ref{H2+H}), despite its astrophysical importance, has received little attention. In their ion cyclotron resonance study at room temperature, Karpas \textit{et al.} \cite{Karpas1979} determined the rate coefficient of all isotopic variants of the reaction, from which they could infer that the reaction does not proceed via scrambling nor via atom transfer, but rather via direct electron transfer. A merged beam study of the H + D$_2^+$ charge tranfer reaction was performed by Andrianarijaona \textit{et al.} \cite{Andrianarijaona2009} down to 1 eV, albeit with hot H$_2^+$ ions as produced by electron impact \cite{vonBusch1972}. Finally, a detailed theoretical study was performed by Krsti\'c \cite{Krstic2002}, who predicts a vibrational distribution of H$_2$ products dominated by $v=4$ down to $\sim 1$ eV, an energy below which all accessible levels become significantly populated.

While reactions (\ref{H2H+}) and (\ref{H2+H}) probe the crossing seam in a full collision, the photodissociation of H$_3^+$ is actually probing it from within, as fragments depart from the classical turning point accessed via a vertical transition from the ground state potential well. In this work, we consider the electronic excitation of the H$_3^+$ ground state to the first excited $^1A'$ state by UV photons of 4 to 5 eV, that triggers rapid dissociation into two or three products:
\begin{eqnarray}
\mathrm{H}_3^+ + h\nu &\rightarrow& \mathrm{H}_2(v) + \mathrm{H}^+ \label{2body1}\\
&\rightarrow& \mathrm{H}_2^+(v^+) + \mathrm{H} \label{2body2}\\
&\rightarrow& \mathrm{H} + \mathrm{H} + \mathrm{H}^+. \label{3body}
\end{eqnarray}
The case of infrared photodissociation of H$_3^+$ to H$_2$ + H$^+$ via quasi-bound resonances at the dissociation limit, as studied by Carrington and Kennedy \cite{Carrington1984}, will not be discussed here, as it does not involve the first excited potential energy surface. 
The adiabatic dissociation limit of the 2 $^1A'$ state is H$_2^+$ + H, while the ground surface dissociates to H$_2$ + H$^+$, the two limits being separated by 1.83 eV in their respective vibrational ground states. The two-body channels H$_2$ + H$^+$ and H$_2^+$ + H are situated 4.48 eV and 2.65 eV below the full atomisation limit, respectively.

A decisive experiment should ideally discriminate between these different channels, measure the associated kinetic energy release and determine the vibrational distribution of the molecular products. Such an experimental effort will be described in the next sections.
The quantum-chemical description of H$_3^+$ allowing for a complete determination of the its potential energy landscape, a time-dependent wavepacket simulation will be shown to give valuable insight into the non adiabatic dynamics associated with the avoided crossing seam.

\section{The avoided crossing seam}

The potential energy surfaces of the H$_3^+$ molecular ion are known to exhibit a rich topology, which results from the high degree of symmetry imposed by the indiscernibility of the protons, and the asymptotic degeneracy of its fragmentation channels that stems from it. We shall discuss it here in terms of Jacobi coordinates $r$, $R$ and $\theta$. In the asymptotic region, $r$ is the internal coordinate of a diatomic fragment, i.e. H$_2$ or H$_2^+$, while $R$ is the distance between the atomic fragment , i.e. H or H$^+$, to the centre-of-mass of the diatom. Numerous calculations of the three-dimensional surfaces exist to this day \cite{Viegas2007,Pavanello2012}, that reveal the existence of an avoided crossing between the ground and first excited potential at distances $R$ > 6 a$_0$. When dealing with two-body breakup, the C$_{\mathrm{2v}}$ symmetry corresponding to $\theta$ = 90$^\circ$, is usually preferred as it allows a simple representation of the surfaces. Its validity for the description of H$_3^+$ dynamics is expected to be somewhat limited, in view of the floppy character of the molecule above its barrier to linearity, located some 10.000 cm$^{-1}$ above its rovibrational ground state \cite{Roehse1994}. 

In their pioneering work, Preston and Tully \cite{Preston1971} had performed a diatomics-in-molecules (DIM) calculation of the ground and first excited $^1A'$ potential energy surfaces of H$_3^+$, revealing the avoided crossing seam resulting from the difference in dissociation energy between H$_2$ and H$_2^+$ exceeding that of ionization potential between H$_2$ and H.  Bauschlicher \textit{et al.} \cite{Bauschlicher1973} have soon after performed the first \textit{ab initio} calculation of those surfaces, and quantified the surface hopping probability at various R distances by reformulating the Landau-Zener-Stueckelberg transition probabilities in terms of adiabatic potentials. Their results were verified and expanded by Ichihara and Yokoyama \cite{Ichihara1995} over the entire domain of Jacobi angle from 0 (collinear) to 90$^\circ$ (isoceles triangle).

\begin{figure}[!h]
\centering\includegraphics[width=3.5in]{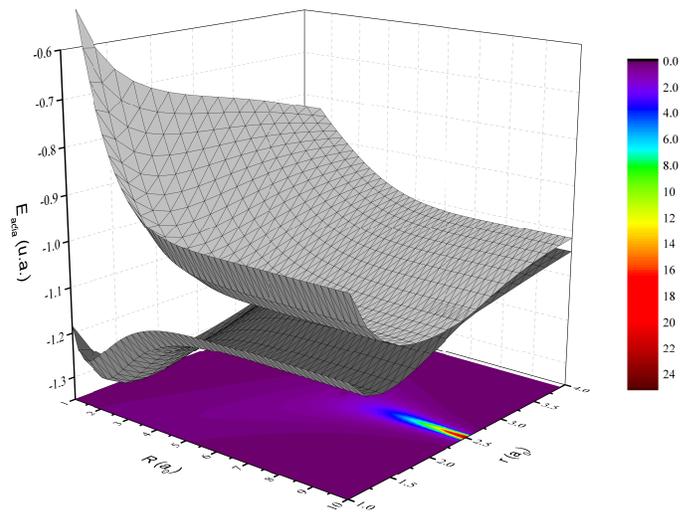}
\caption{Adiabatic potential energy surfaces of the two first $^1A_1$ states of H$_3^+$. Colour map: non-adiabatic coupling matrix elements (a.u.) along $r$.}
\label{fig_adiab}
\end{figure}

In order to be able to run dynamical simulations, we have generated these potential energy surfaces in $C_{2v}$ with the CASSCF + Full CI method as implemented in \texttt{MOLPRO} \cite{Werner2012}, the results of which are shown in figure \ref{fig_adiab}. The energies were calculated on a grid of 150 points in $r$ (from 0.6 to 38 a$_0$ with additional points close to the crossing) and 69 points in $R$ (from 1 to 20 a$_0$) with a total of 10350 points. The basis set is AVTZ supplemented by additional functions, as employed for HeH$^+$ by Loreau \textit{et al.} \cite{Loreau2010}. The non-adiabatic coupling matrix elements (NACME) were calculated on the same grid as the PES with the three-point method. Their value is represented in the colour map under the surface plot. The cusp corresponding to the avoided crossing appears at $r$ $\simeq$ 2.5 a$_0$ for distances $R$ beyond 6 a$_0$. The non-adiabatic coupling is in good agreement with previous calculations \cite{Barragan2004,Barragan2006}.

\section{Ultraviolet photodissociation of H$_3^+$}

Photodissociation, being essentially a half-collision process, offers a convenient probe of the long-range dynamics without the impact parameter averaging imposed by full collisions. 
In their seminal work on time-dependent wavepacket propagation, Kulander and Heller \cite{Kulander1978} have computed the photodissociation cross section for H$_3^+$ in its absolute ground state. The cross section was found to peak around 21 eV, rendering the photodissociation of H$_3^+$ in astrophysical environments rather unlikely \cite{vanDishoeck1987}. The calculation included the first  $^1B_2$ potential energy surface (in $C_{2v}$ symmetry), which is degenerate with the 2 $^1A_1$ at equilibrium geometry. While the dissociation along that third surface exclusively feeds the three-body channel H +H + H$^+$, the ground and first excited $^1A_1$ surfaces were treated in a diabatic picture, neglecting the interaction between them. The vibrational population of the H$_2^+$ products was found to peak at $v^+$ = 0, a result we shall put in perspective with our experimental findings.

Such photodissociation events were recorded by Bae and Cosby \cite{Bae1990} in a fast beam experiment in collinear geometry where H$_2^+$ fragments were electrostatically separated and counted. Not surprisingly, the process was found to be absent at photon energies below 2.5 eV, while its yield increased steadily with photon energy above that. An apparent cross section as low as $10^{-22}$ cm$^2$ was measured at 4 eV, which is five orders of magnitude below the photodissociation cross section of H$_2^+$.

\begin{figure}[!h]
\centering\includegraphics[width=3.2in]{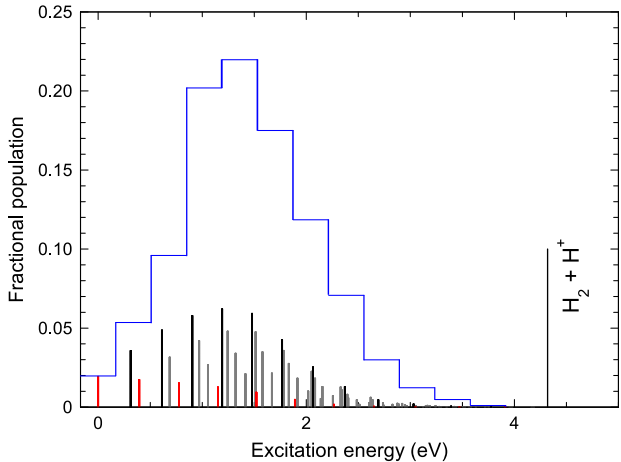}
\caption{Statistical population of ($\nu_A$, $\nu_E$) modes (red: ($\nu_A$,0) levels, black: (0,$\nu_E$) levels, grey: combined excitations) following H$_3^+$ formation through reaction (\ref{H3formation}) from \cite{Anicich1984}. Histogram: binned excitation energy distribution.}
\label{fig_popH3}
\end{figure}

This observation may be rationalised with the help of the internal energy distribution resulting from the H$_3^+$ formation process:
\begin{equation} \label{H3formation}
\mathrm{H}_2 + \mathrm{H}_2^+ (v^+) \rightarrow \mathrm{H}_3^+(\nu_A,\nu_E) + \mathrm{H}.
\end{equation}
Anicich and Futrell \cite{Anicich1984} have calculated the population of the symmetric ($\nu_A$) and doubly degenerate asymmetric ($\nu_E$) mode for reaction (\ref{H3formation}) with the help of a statistical model accounting for the vibrational distribution of the H$_2^+$ reactant, as depicted in figure \ref{fig_popH3}. One observes a tail in the distribution that extends towards the dissociation limit. Bae and Cosby however operated with a source consisting of a pulsed valve coupled with an electron gun, causing the nascent internal excitation to be substantially quenched during the expansion of the gas plume.

More recently, Alexander \textit{et al.} \cite{Alexander2009} have observed the decay of the photodissociation signal in an ion beam trap. D$_3^+$ ions were produced in an electron cyclotron resonance source and stored between electrostatic mirrors. An intense femtosecond laser operating at 800 nm  caused photodissociation of a decaying fraction of the stored ions as monitored by the detection of neutral particles leaving the trap. At such a long wavelength, the process is multiphotonic in nature, the molecular ions likely absorbing up to three photons at the highest intensities used in the experiment. More complete experiments were performed at 790 nm by Sayler \textit{et al.} \cite{Sayler2012}.

Petrignani \textit{et al.} \cite{Petrignani2010} have reported a similar experiment performed at the TSR ion storage ring in Heidelberg with a nanosecond laser of moderate intensity operating on the second and fourth harmonic of the Nd:YAG laser, i.e. 532 nm and 266 nm, respectively. In this single photon regime, the photodissociation signal, as observed by the detection of H$_2^+$ ions leaving the storage ring orbit, exhibit a rapid decay in the millisecond range, that was attributed to the depopulation of the tail of the distribution of internal states by radiative cooling. Complementary measurements were performed with a single-pass set-up in Louvain-la-Neuve, which will be described in some more detail below since new measurements have been performed to corroborate theoretical findings. 

\subsection{Experimental set-up}

\begin{figure*}
\centering\includegraphics[width=5 in]{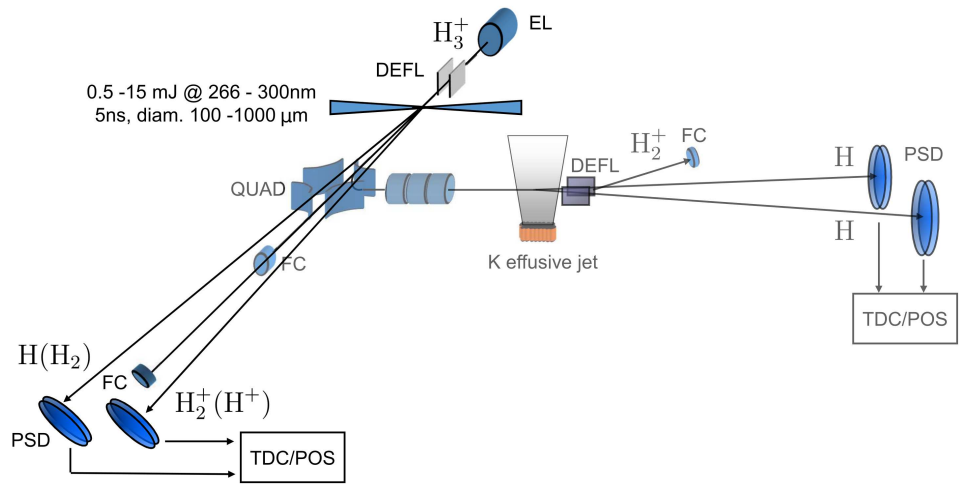}
\caption{Experimental set-up. EL: Einzel lens, DEFL: electrostatic deflector used as beam chopper, QUAD: quadrupole deflector, FC: Faraday cup, PSD: position-sensitive detectors, TDC: time-to-digital converter, POS: position analyser.}
\label{fig_photodis2}
\end{figure*}

Our experimental setup is based on a small scale accelerator delivering beams of tens of nanoamperes at a few keV. The ions are created in a duoplasmatron source by electron impact ionization of H$_2$ and subsequent collisions, and should adopt an internal energy distribution similar to what was predicted by Anicich and Futrell \cite{Anicich1984} (figure \ref{fig_popH3}). After acceleration, the beam is strongly collimated and crossed at right angle with the pulsed laser beam (figure \ref{fig_photodis2}) 
Experiments have been conducted at two different wavelengths, i.e. 300 nm and 266 nm. The former was produced by a dye laser pumped by the second harmonic of a nanosecond Nd:YAG laser (Continuum), while the latter is the fourth harmonic of the Nd:YAG obtained by frequency doubling the second harmonic output of the Nd:YAG integrated in our OPO laser system (Ekspla). Both lasers operate at a repetition rate of 30 Hz. The ion beam is chopped in 500 ns-long bunches synchronised to the laser shots in order to limit the load on the detectors located 1.75 m downstream. The primary H$_3^+$ ions are collected in a small Faraday cup located in front of the pair of detectors consisting of a Z-stack of microchannel plates backed with a resistive anode (Quantar).

The detectors are arranged as to minimize the dead area between them. To this goal, the second detector has been placed 10 cm downstream of the first one. This offers a second advantage: ions passing along the first detector's back plane get deflected sideways, allowing their easy separation from the neutrals impinging the second detector. The selection of a particular event is based on the successful reconstruction of its centre-of-mass both in time and position, i.e. the arrival time and location of the non-dissociated H$_3^+$. Note however that this two-detector arrangement does not allow for the detection of the three-body breakup, reaction (\ref{3body}). 

Since the vibrational excitation cannot be unequivocally determined from the measured kinetic energy release due to the unknown starting point in the H$_3^+$ potential well, one must rely on a post-interaction analysis of the products. The latter is performed by means of our dissociative charge transfer method as described in our study of the vibrational distribution of H$_2^+$ resulting from intense laser-field ionisation of H$_2$  \cite{Urbain2004}. In a nutshell, the molecular ions are deflected by an electrostatic quadrupole towards an effusive potassium jet, where they undergo resonant electron capture to the $a$ $^3\Sigma_g^+$ and $c$ $^3\Pi_u$ states of H$_2$. The former quickly radiates to the repulsive $b$ $^3\Sigma_u^+$  state while the latter is rotationally predissociated by the same $b$ state, resulting in a pair of ground state atoms whose kinetic energy bears the imprint of the initial vibrational excitation \cite{DeBruijn1984}. A last complication arises from the fact that the molecular ions themselves have a kinetic energy that depends on the initial photodissociation event. Time-of-flight selection applied to the centre-of-mass of the two hydrogen atoms allows for a complete reconstruction of the stepwise fragmentation. Note that the same recipe does not apply to the vibrational analysis of H$_2$ products.

\subsection{Experimental results}

\begin{figure} [!h]
\centering\includegraphics[width=3.2 in]{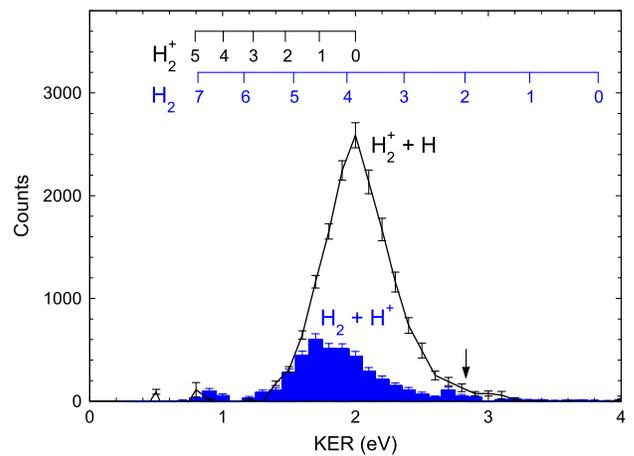}
\caption{Measured kinetic energy release distributions for the H$_2^+$ + H and H$_2$ + H$^+$ channels.  The arrow marks the expected KER for H$_3^+$ at the dissociation limit. Integrated counts reflect the actual branching ratio.}
\label{fig_H3_266}
\end{figure}

Kinetic energy release (KER) measurements have been performed under varying ion source conditions, and over a wide range of laser intensities, in order to investigate the role of initial excitation of the ions, and the possible occurrence of secondary photodissociation of the H$_2^+$ products. The KER distributions resulting from photodissociation of H$_3^+$ at 266 nm are shown in figure \ref{fig_H3_266}. Vibrational ladders are placed assuming the peak of the H$_2^+$ + H distribution coincides with $v^+$ = 0, as discussed below.

Similar results were obtained at 300 nm, although the ground state channel (reaction (\ref{2body1})) was most often barely visible amongst the excessive collisional background present in our first set of experiments. In both cases, the distribution of H$_2^+$ + H events (reaction (\ref{2body2})) adopts a triangular shape, centred around 2 and 1.67 eV at 266 nm and 300 nm, respectively, while the ground state channel, H$_2$ + H$^+$, peaks at slightly lower energy, i.e. 1.75 eV at 266 nm. The branching ratio is about 4:1 in favour of the H$_2^+$ + H channel (see table \ref{table_vib}), confirming the intervention of the crossing seam in the dissociation dynamics.  

Our KER spectra are remarkably similar to the measurements of Gaire \textit{et al.} \cite{Gaire2012} performed with 40 fs pulses at 395 nm. They also observed a 0.25 eV shift of the D$_2$ + D$^+$ peak towards lower energy, and a branching ratio of about 6:1 between the two channels. Considering the high intensity involved ($10^{14}$ W cm$^{-2}$), they assigned the D$_2$ + D$^+$ and D$_2^+$ + D contributions to 2- and 3-photon processes, respectively. The present results challenge this interpretation. 

\begin{table*}
\caption{Fractional population (in percent) of vibrational levels of H$_2^+$ measured at $\lambda$ = 266 and 300 nm (exp). Two-dimensional wavepacket calculations at 58 nm (Kulander and Heller \cite{Kulander1978}), 266 nm and 300 nm (WP, present work). Vibrational distribution of H$_2$ from present wavepacket calculations. BF: branching fraction to H$_2^+$ + H or H$_2$ + H$^+$.}
\label{table_vib}
\begin{tabular}{cccccccccc}
\hline
& 58 nm & &\multicolumn {3}{c}{266 nm} & &\multicolumn {3}{c}{300 nm} \\
v & H$_2^+$ \cite{Kulander1978} & \ & H$_2^+$ (exp)& H$_2^+$ (WP) & H$_2$ (WP) & \ &  H$_2^+$ (exp) & H$_2^+$ (WP) &  H$_2$ (WP)\\
\hline
0 & 75 & & 72(4) & 88.0 & 0.0 & & 63(5) & 89.9 & 0.0\\
1 & 19 & & 20(4) & 3.1 & 0.0 & & 23(5) & 1.6 & 0.0\\
2 & 1.5 & & 6(4) & 5.6 & 0.0 & & 13(5) & 6.0 & 0.3\\
3 & 2.5 & & 2(4) & 3.0 & 1.9 & & 1(5) & 2.2 & 1.2\\
4 & 1.6 & & & 0.3 & 48.7 & & & 0.3 & 55.5\\
5 & 0.4 & & & 0.1 & 46.1 & & & 0.0 & 41.5\\
6 & 0.0 & & & & 3.2 & & & & 1.5\\
7 & & & & & 0.0 & & & & 0.0\\
\hline
BF & 100 & & 79(4) & 70.9 & 29.1 & & 87(7) & 71.5 &28.5\\
\hline
\end{tabular}
\end{table*}

\begin{figure}[!h]
\centering\includegraphics[width=3.2in]{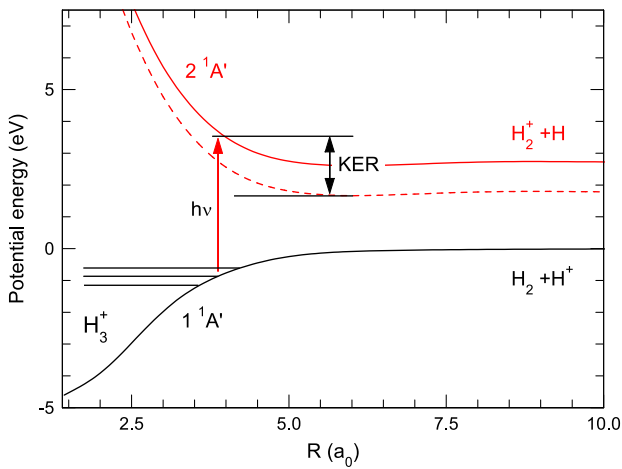}
\caption{Schematic representation of the UV excitation from the 1 $^1A'$ to the 2 $^1A'$ state. Full (dashed) lines are cuts through the potential energy surfaces at the equilibrium distance of H$_2$ (H$_2^+$). The actual kinetic energy release depends on the vibrational excitation of the H$_2^+$ product.}
\label{fig_photodis1}
\end{figure}

In order to confirm the dominant population of $v^+$ = 0, we also measured the vibrational distribution of emerging H$_2^+$ ions by means of our dissociative charge transfer method. For that purpose, the intensity of the H$_3^+$ beam had to be increased to hundreds of nA, as the combined charge transfer and coincidence detection efficiency does not exceed 10$^{-3}$. The analysis of the vibrational distributions is summarised in table \ref{table_vib}. More than two thirds of the population is concentrated in the $v^+=0$ level, while no population could be detected beyond $v^+=3$. This finding matches astonishingly well the theoretical prediction of Kulander and Heller \cite{Kulander1978}, which is quite surprising when considering that they computed the distribution for cold H$_3^+$ while we measured with fairly hot ions. Moreover, their calculation deliberately ignored the presence of the avoided crossing seam, which is likely to contribute to ground state dissociation. 

Building on this knowledge, one may locate the energy levels contributing the most to the photodissociation signal by subtracting the KER and the energy difference between the H$_2^+$ +H and H$_2$ + H$^+$ asymptotes to the photon energy, as depicted in figure \ref{fig_photodis1}. These levels would sit 0.83 eV and 0.6 eV below the H$_3^+$ dissociation limit, at 266nm and 300 nm respectively, which according to the work of Anicich and Futrell \cite{Anicich1984}, constitute $\sim 0.5$ \% of the nascent vibrational distribution (see fig. \ref{fig_popH3}). This tiny population is compatible with the apparent photodissociation cross section measured at 266 nm by Petrignani \textit{et al.} \cite{Petrignani2010} , i.e. $\sim 7 \times 10^{-20}$ cm$^2$.

\subsection{Wavepacket simulations}

In order to interpret these experimental observations, we performed time-dependent wavepacket calculations on the first two coupled potential energy surfaces. 
The time-dependent approach relies on the projection of the initial rovibrational wavefunction on the upper surface, after multiplication by the dipole matrix element.
A more complete description of the photodissociation would in principle require to include the second excited singlet potential energy surface, which presents a conical intersection with the first excited potential in equilateral configurations \cite{Viegas2007}. However, our simulations show that at the wavelengths explored here, the wave packet does not reach this region.

\begin{figure} [!h]
\centering\includegraphics[width=3.3 in]{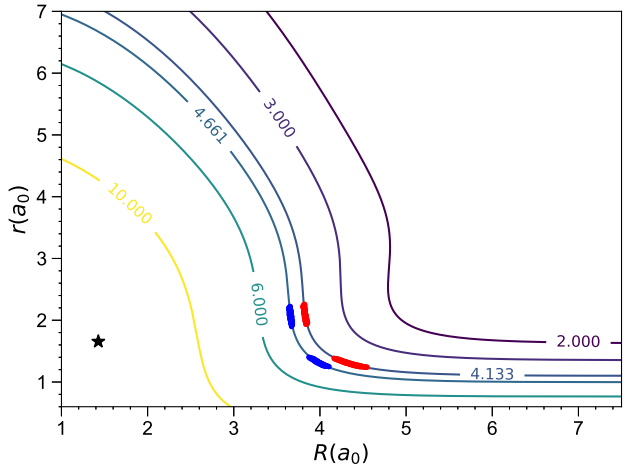}
\caption{Contour lines marking the region where the separation between the ground and first excited potential energy surfaces matches the indicated photon energies. The most likely transition regions at 266 nm (4.661 eV, blue) and 300 nm (4.133 eV, red) are marked by bold lines. The star marks the equilibrium geometry of H$_3^+$.}
\label{fig_contour}
\end{figure}

As shown in figure \ref{fig_popH3}, many vibrational states of H$_3^+$ are populated in the experiment. 
To identify the geometry where the initial wavepacket is defined on the excited potential energy surface, one may use the information provided by the KER and follow the principle illustrated by figure \ref{fig_photodis1}: the vertical transition from the ground to the first excited $^1A'$ surface must occur along a path corresponding to the so-called Condon point, where the potential energy difference matches the photon energy. Such contour lines are represented on figure \ref{fig_contour} in $C_{2v}$ symmetry for photon energies ranging from 2 to 10 eV. The location of these transition points on the upper surface for the photon energies in the present experiment (4.133 eV or 4.661 eV) spans a wide energy domain (see figure \ref{fig_adiab}), which in turn would generate a broad kinetic energy distribution, as shown by our experimental results (figure \ref{fig_H3_266}). Added in bold on figure \ref{fig_contour} are the regions contributing to the peak of the KER distribution when considering H$_2^+(v^+=0)$ products only.

To assess the role of the avoided crossing seam, time-dependent wavepackets were propagated on diabatic surfaces derived from our adiabatic potential energy surfaces and NACME (figure \ref{fig_adiab}). A strict diabatisation was performed along the $r$ coordinate with $R$ fixed, since the avoided crossing is along the $r$ coordinate. The two-dimensional wavepacket propagation was performed with the computer package \textsc{WavePacket} of Schmidt and Lorenz \cite{Schmidt2017}. Gaussian wavepackets were launched from a selection of starting coordinates (marked in bold on figure \ref{fig_contour}). Their vibrational analysis follows the method of Balint-Kurti \textit{et al.} \cite{BalintKurti1990}. The Fourier transform of the wavepacket $\Phi(R,r,t)$ is computed at sufficiently large $R_\infty$ and projected onto the vibrational wavefunctions $\chi_v(r)$ of the diatomic (H$_2$ or H$_2^+$):
\begin{equation}
A_v(R_\infty,E)=\frac{1}{2\pi} \int_0^\infty \exp(iEt/\hbar)\langle\chi_v(r)|\Phi(R_\infty,r,t)\rangle dt.
\end{equation}
The square of the $A_v$ amplitude is identified as the population of vibrational level $v$. The choice of the total energy $E$ is somewhat arbitrary as several H$_3^+$ levels may contribute to the photodissociation signal at a given wavelength. For consistency, it is taken equal to the potential energy of the 2 $^1A_1$ state at the initial position of the wavepacket. 

The vibrational populations obtained when launching the wavepacket at ($R$ = 3.67 a$_0$, $r$ = 2 a$_0$) for 266 nm and ($R$ = 3.8 a$_0$, $r$ = 2 a$_0$) for 300 nm
are given in table \ref{table_vib}. Both vibrational distributions are in fair agreement with experiment, with the H$_2^+$ population peaking at $v^+=0$ while the H$_2$ population is concentrated in $v=4$ and 5, as suggested by the vibrational ladder in figure \ref{fig_H3_266}. Moreover, the computed branching ratio 0.71:0.29, is not too distant from the experimental values. 
Wave packet simulations were performed for initial geometries close to those reported above, indicated in Fig. 6, with qualitatively similar conclusions regarding the distribution of final vibrational states.

While these encouraging results confirm our interpretation, the  initial state of H$_3^+$ is still ill-defined, borrowing from our knowledge of the average kinetic energy release. Specific rovibrational wavefunctions should in principle serve as an input, which is beyond the applicability of our model due to its reduced dimensionality.
Obviously, a multidimensional treatment of the non-adiabatic interactions is also in order, as was recently performed by Alijah \textit{et al.} \cite{Alijah2015} and Mukherjee \textit{et al.} \cite{Mukherjee2016}.
Such a full-dimensional quantum mechanical treatment was reported by Sun \textit{et al.} \cite{Sun2015}. While the authors present a kinetic energy spectrum for the photodissociation of H$_3^+$ at 266 nm, no mention is made of the avoided crossing seam, which is somewhat surprising considering the amount of computational effort.

\section{Conclusion}

We have carried out an experimental study of the UV photodissociation of H$_3^+$ towards H$_2$ + H$^+$ and H$_2^+$ + H channels at wavelengths of 266 nm and 300 nm. Such wavelengths mainly probe the photodissociation from excited vibrational states of H$_3^+$ that are populated following its formation in the ion source plasma. The photodissociation is known to proceed through two potential energy surfaces that are connected by a seam of non-adiabatic couplings. While this seam plays a minor role in the photodissociation from the ground vibrational state of H$_3^+$, the present measurements indicate that this is not the case for vibrationally excited ions. An analysis of the kinetic energy release of both channels was performed, showing a dominant contribution from the H$_2^+$ + H channel with H$_2^+$ in the ground vibrational state. On the other hand, for the H$_2$ fragments a strong vibrational excitation was observed. 
These findings were rationalised by performing time-dependent wavepacket simulations of the photodissociation process, demonstrating the impact of the non-adiabatic seam on the dynamics. While the theoretical model was limited by considering only two reaction coordinates as well as by the uncertainty on the initial vibrational distribution of H$_3^+$, it nevertheless provided a clear qualitative interpretation of the experimental results.

A natural extension of the present work would be to include the $^1B_2$ potential energy surface that is degenerate with the 2 $^1A_1$ in equilateral triangle geometry. More importantly,  one must consider different values of the Jacobi angle, since the latter was shown to be a crucial parameter controlling the vibrational excitation of H$_2$ in reaction (\ref{H2H+}) \cite{Errea2007}.

The photodissociation may not be of immediate astrophysical relevance, as it is listed in UMIST and other astrophysical databases with a negligible rate of $5\times 10^{-15}$s$^{-1}$.This is solely due to the absorption window being located around 21 eV. Assuming there are environments where vibrationally excited H$_3^+$ are exposed to VUV radiation, a moderate vibrational excitation could already lead to a substantial shift of the absorption window towards longer wavelengths due to the steepness of the upper potential energy surface.

The knowledge gathered in the present study is readily transposable to the charge transfer reactions (\ref{H2+H}) and (\ref{H2H+}), as they share the same avoided crossing seam dynamics. In that respect, the study of photodissociation at longer wavelength would give access to vibrational states of large deformation as predicted to exist close to the dissociation threshold \cite{Munro2015}. Such studies will serve as an additional benchmark of the existing surfaces and non-adiabatic coupling matrix elements.
\vskip6pt

\enlargethispage{20pt}

\acknowledgments
The authors thank H. Kreckel for fruitful discussions. They are indebted to J. Li\'evin for having suggested to investigate the role of the crossing seam motivating the present study.

This work was supported by the Fonds de la Recherche Scientifique-FNRS through IISN grant No. 4.4504.10. Computational resources have been provided by the Consortium des Équipements de Calcul Intensif (C\'ECI), funded by the Fonds de la Recherche Scientifique-FNRS under Grant No. 2.5020.11. The authors thank the Belgian State for the grant allocated by Royal Decree for research in the domain of controlled nuclear fusion. XU is Senior Research Associate of the Fonds de la Recherche Scientifique-FNRS.

\end{document}